\documentclass[aps,showpacs,twocolumn,floatfix,amsmath,preprintnumbers,nofootinbibe,secnumarabic]{revtex4} 

\usepackage{amsmath,amssymb,hyperref,url}

\def\art{paper}

\def\boo#1#2#3#4#5#6#7{#1. \textit{#2}; #3: #4, #5, #6.} \def\jrn#1#2#3#4#5#6{#1. #2. \textit{#3} \textbf{#6}, \textit{#4}, #5.} \def\andd{.; } \def\Ref{} \def\Refs{}  \def\eq{Equation } \def\eqs{Equations } \def\eqref#1{(\ref{#1})} \def\Sec{Section} \def\Secs{Sections } 

\def\Ref{Ref.} \def\Refs{Refs.} \def\eq{Eq. } \def\eqs{Eqs. }  

\def\citen#1{\cite{#1}}
\def\scn#1#2{\section{#1}\lb{#2}} \def\sscn#1#2{\subsection{#1}\lb{#2}}

\def\bfl{\begin{flushleft}}
\def\efl{\end{flushleft}}
\def\bfr{\begin{flushright}}
\def\efr{\end{flushright}}
\def\bc{\begin{center}}
\def\ec{\end{center}}
\def\be{\begin{equation}}
\def\ee{\end{equation}}
\def\bse{\begin{subequations}}
\def\ese{\end{subequations}}
\def\ba{\begin{eqnarray}}
\def\ea{\end{eqnarray}}
\def\baa#1{\begin{array}{#1}}
\def\eaa{\end{array}}
\def\bw{} \def\bw{\begin{widetext}}
\def\ew{} \def\ew{\end{widetext}}
\def\nn{\nonumber }
\def\lb#1{\label{#1}}
\def\bit{\begin{itemize}}
\def\eit{\end{itemize}}
\def\bco{}
\def\bcs{\begin{cases}}
\def\ecs{\end{cases}}

\def\Sign#1{\, \text{sign}\left(#1\right) }

\def\Der#1#2{\frac{\drm #1}{\drm #2}}

\def\Sin#1#2{\, \text{sin}^{#1}#2}

\def\Kummer#1#2#3{\,_1F_1\!\left(#1, #2; #3 \right)}
\def\Auxfun#1#2{\, \breve{Z}\left(#1, #2 \right)}
\def\Prlog#1{\,{\cal W}\!\left(#1 \right)}

\def\cf{\eta}

\def\vol{{\cal V}}

\def\Mass{{\cal M}}
\def\lan{{\cal L}}
\def\lanp{{V}}

\def\vena{\boldsymbol{\nabla}}
\def\ve#1{\boldsymbol{#1}}

\def\drm{d}
\def\dvol{\drm\vol}

\def\dn{\rho}  \def\dnc{\bar{\dn}}
\def\lnc{\bar{\ell}}

\def\tfc{q}
\def\lapl{\vena^2}

\def\nc0{b_0}

\def\tps{T_\Psi}

\def\wfv{\Psi_{\text{vac}}}
\def\wfvo{\Psi_{(0)}}

\def\grpot{\Phi}
\def\grpott{\grpot} 
\def\grsmi{\grpot_{\text{smi}}}
\def\grmic{\grpot_{\text{RN}}}
\def\grnew{\grpot_{\text{N}}}
\def\grgal{\grpot_{\text{gal}}}
\def\grmgal{\grpot_{\text{mgl}}}
\def\grcos{\grpot_{\text{dS}}}

\def\lsmi{L_{\text{smi}}}
\def\lmic{L_{\text{RN}}}

\def\lmgl{L_{\text{mgl}}}
\def\lcos{L_{\text{dS}}}
\def\lpwr{L_\pwr}

\def\rcos{R_{\text{dS}}}

\def\rhcos{\rho_{\text{dS}}}

\def\ssmgl{{\virgo}}
\def\ssmgl{{\text{mgl}}}
\def\ssmgl{{M}}
\def\ssmgl{{\mathfrak M}}

\def\rad{r}
\def\rve{\ve{r}}
\def\pwr{\chi}
\def\dil{{\phi}}
\def\ror{R}
\def\dnind{\rho_\grpott}
\def\dnir{\tilde{\rho}}
\def\dnindr{{\dnir}_\grpott}
\def\adn#1{\widetilde{\Omega}_{\text{#1}}}

\def\mesph{\sigma^2} 

\begin{document}




\preprint{\scriptsize \it
17th Russian Gravitational Conference - International Conference on Gravitation, Cosmology and Astrophysics (RUSGRAV-17)~}

\preprint{\footnotesize Universe \textbf{6}, 180 (2020)   
\ \ 
[\href{https://doi.org/10.3390/universe6100180}{DOI: 10.3390/universe6100180}]
}

\title{ ~\\
An alternative to dark matter and dark energy: 
Scale-dependent gravity\\ in superfluid vacuum theory}
\author{Konstantin G. Zloshchastiev}
\email{https://orcid.org/0000-0002-9960-2874} 
\affiliation{Institute of Systems Science, Durban University of Technology, P.O. Box 1334, Durban 4000, South Africa}


\begin{abstract}
We derive an effective gravitational potential, 
induced by the quantum wavefunction of a physical vacuum of a self-gravitating configuration, while the vacuum itself is viewed as the superfluid described by the logarithmic quantum wave equation. We determine that gravity has a multiple-scale pattern, to such an extent that one can distinguish sub-Newtonian, Newtonian, galactic, extragalactic and cosmological terms. The last of these dominates at the largest length scale of the model, where superfluid vacuum induces an asymptotically Friedmann-Lema\^itre-Robertson-Walker-type spacetime, which provides an explanation for the accelerating expansion of the Universe.
The model describes different types of expansion mechanisms, which could explain the discrepancy between measurements of the Hubble constant using different methods. 
On a galactic scale, our model explains the non-Keplerian behaviour of galactic rotation curves, and also why their profiles can vary depending on the galaxy. It also makes a number of predictions about the behaviour of gravity at larger galactic and extragalactic scales. We demonstrate how the behaviour of rotation curves varies with  distance from a gravitating center, growing from an inner galactic scale towards a metagalactic scale: a squared orbital velocity's profile crosses over from Keplerian to flat, and then to non-flat. The asymptotic non-flat regime is thus expected to be seen in the outer regions of large spiral galaxies.
\end{abstract}


\date{received: 
17 June 2020 [RG], 29 August 2020 [MDPI], 26 November 2020 [arXiv]}


\maketitle

\scn{Introduction}{s:intro}
Astronomical observations over many length scales
support the existence of a number of novel phenomena, which are usually attributed
to dark matter (DM) and dark energy (DE).
Dark matter was introduced to explain a range of observed phenomena
at a galactic scale, such as flat rotation curves,
while 
dark energy 
is expected to account for cosmological-scale dynamics,
such as the accelerating expansion  of the Universe.
For instance, the~$\Lambda$CDM model, which is currently the most popular approach 
used 
in cosmology and galaxy-scale astrophysics,
makes use of both DE and cold DM concepts~\cite{car01}.
In spite of being a generally successful framework purporting to explain the large-scale structure of the Universe,
it currently faces certain challenges ~\cite{bb17,ty18}.

There is also growing consensus that a convincing theory of DM- and DE-attributed phenomena cannot 
be a stand-alone model; but should, instead, be a part of a fundamental theory involving all known interactions.
In turn, we contend that formulating this fundamental theory will be impossible without
a clear understanding of the dynamical structure of the physical vacuum,
which underlies all interactions that we know of.
Moreover, this theory must operate at a quantum level, which~necessitates us rethinking
of the concept of gravity using basic notions of quantum~mechanics.

One of the promising candidates for a theory of physical vacuum is superfluid vacuum theory (SVT),
a post-relativistic approach to high-energy physics and gravity.
Historically,
it evolved from Dirac's idea of 
viewing the physical vacuum as a nontrivial quantum object,
whose phase and derived velocity are non-observable in a {quantum-mechanical sense}
~\cite{dir51}.
The term `post-relativistic', 
in this context,
means that SVT can generally be a non-relativistic theory;
which nevertheless contains relativity 
as a special case, or~limit, with~respect to some dynamical value 
such as momentum 
(akin to general relativity being a superset of the Newton's theory of gravity).
Therefore, underlying three-dimensional space would not be physically observable
until an observer goes beyond the above-mentioned limit, as~will be discussed in more detail later in this~article.

The dynamics and structure of superfluid vacuum are being studied, using various approaches 
which agree upon the main paradigm (physical vacuum being a background quantum liquid of a certain kind,
and elementary particles being excitations thereof),
but differ in their physical details, such as an underlying model of the
liquid~\cite{volbook,huabook,z11appb}. 

It is important to work with a precise definition of superfluid,
to ensure that we avoid the most common misconceptions which otherwise might arise when one attempts to apply superfluid models to astrophysics and cosmology, some details can be found in Appendix \ref{s:sf}.
In fact, some superfluid-like models of dark matter based on classical perfect fluids, scalar field theories or scalar-tensor gravities,
turned out to be vulnerable to experimental verification~\cite{lmo19}.
Moreover, superfluids are often confused not only with perfect fluids,
but also
with 
the concomitant phenomenon of 
Bose-Einstein condensates (BEC),
which is
another kind of quantum matter occurring in low-temperature condensed matter
~\cite{psbook04}.
However, 
even though BEC's do share certain features with superfluids, 
this does not imply that they are 
superfluidic in~general.

In particular, quantum excitations in laboratory superfluids that we know of 
have dispersion relations of a distinctive shape
called the Landau ``roton'' spectrum.
Such a shape of the spectral curve is crucial, as~it ensures 
the suppression of dissipative fluctuations at { a quantum level}
~\cite{z12eb,z20ijmpa},
which~results in inviscid flow~\cite{kap38,am38}.
If plotted as an excitation energy versus momentum,
the curve starts from the origin, climbs up to a local maximum (called the {maxon} peak), 
then slightly descends to a local nontrivial minimum (called the ``roton'' energy gap); then grows  again, this time all the way up, 
to the boundary of the theory's applicability range.
In fact it is not the roton energy gap alone, but~the energy barrier formed by the maxon peak and roton minimum in momentum space,
which~ensures the above-mentioned suppression of quantum fluctuations in quantum liquid
and, ultimately, causes its flow to become inviscid.
In other words, it is the global characteristics of the dispersion curve, not just the existence of a nontrivial
local minimum and related energy gap,
which is important for superfluidity to occur. 
Obviously, these are non-trivial properties,
which cannot possibly occur in all quantum liquids and condensates.
Further details and aspects are discussed in Appendix \ref{s:sf}.


This \art~ is organized as follows.
Theory of physical vacuum based on the logarithmic superfluid model
is outlined in \Sec~\ref{s:mod},
where we also demonstrate how four-dimensional spacetime can emerge from the
three-dimensional dynamics of quantum liquid. 
In \Sec~\ref{s:gr}, we derive the gravitational potential,
induced by the logarithmic superfluid vacuum in a given state,
using certain simplifying assumptions.
Thereafter, in~\Sec~\ref{s:phy},  
we give a brief physical interpretation of different parts of the derived 
gravitational potential and estimate their characteristic length scales.
In \Sec~\ref{s:den},
profiles of induced matter density 
are derived and discussed for the case of spherical symmetry. 
Galactic scale phenomena are discussed 
in \Sec~\ref{s:frc}, 
where
the phenomenon of galactic rotation curves is explained 
without introducing any exotic matter ad~hoc.
In \Sec~\ref{s:uexp},
we discuss the various mechanisms of the accelerating expansion of the Universe, 
as well as the
cosmological singularity, 
``vacuum catastrophe'' 
and
cosmological coincidence
problems. 
Conclusions are drawn in \Sec~\ref{s:con}.\\

\scn{Logarithmic Superfluid Vacuum}{s:mod}
Superfluid vacuum theory assumes that 
the physical vacuum is described,
when disregarding
quantum fluctuations, by~
the
fluid
condensate wavefunction $\Psi (\rve,\, t)$,
which is a three-dimensional Euclidean scalar.
The state itself is described by a ray in the corresponding Hilbert space,
therefore
this wavefunction obeys a normalization condition
\be\lb{e:norm}
\langle \Psi | \Psi \rangle
= \int_\vol \dn \, \dvol 
= \Mass
,\ee 
where 
$\Mass$ and $\vol$  
are the
total mass and volume of the fluid, respectively,
and
$\dn = |\Psi|^2$ is the fluid mass density.
The wavefunction's dynamics is governed by an equation of  
a $U(1)$-symmetric Schr\"odinger~form:
\be\lb{e:becgeneq}
\left[
- i \hbar \, \partial_t
- \frac{\hbar^2}{2 m} \vena^2
+
V_{\text{ext}} (\rve,\, t)
+
F(|\Psi|^2)
\right]
\Psi
= 0,
\ee
where 
$m$ 
is the constituent particles' 
mass,
$V_\text{ext} (\rve,t)$ is an external or trapping potential
and $F (\dn)$ is a duly chosen function, which effectively 
takes into account many-body effects inside the fluid.
This~wave equation can be formally derived as a minimizing condition of an action functional
with the following
Lagrangian:
\bw
\ba
\lan
&=& 
\frac{i \hbar}{2}(\Psi \partial_t\Psi^* - \Psi^*\partial_t\Psi)+
\frac{\hbar^2}{2 m}
|\vena \Psi|^2
+ V_\text{ext} (\rve,\, t)\, 
|\Psi|^2
+
\lanp (|\Psi|^2)
,
\lb{e:ftlan}
\ea
\ew
where $\lanp (\dn)$ equals to a primitive of $F(\dn)$ up to an additive constant:
$F (\dn) = \lanp' (\dn) $;
throughout the paper the prime denotes a derivative with respect to the function's~argument.

In this picture, 
massless excitations, such as photons, are analogous to  
acoustic waves propagating with velocity 
$ 
 c_s \propto \sqrt{|p' (\dn) |}
$, 
where fluid pressure $p =p (\dn)$ is determined via the equation of state.
For~the system \eqref{e:becgeneq},
both the equation of state and speed of sound can be derived using
the fluid-Schr\"odinger analogy, which was established for a special case 
in \Ref~\cite{dmf03},
and generalized for an arbitrary $F (\rho)$  in works~\cite{z11appb,z19mat}.
In a leading-order approximation
with respect
to the Planck constant, 
we obtain
\be\lb{e:cappapp}
p 
= 
- \frac{1}{m} 
\int\!\rho F'(\rho)\, \drm\rho
,
\ \
c_s^2
= 
\frac{1}{m} 
\dn
|F' (\dn)|
,
\ee
while higher-order corrections 
would 
induce Korteweg-type effects, thus significantly
complicating the subject matter~\cite{z19mat}.

Furthermore,
it is natural to require that
superfluid vacuum theory must 
recover Einstein's theory of relativity at a certain limit. 
One can show that at a limit of low momenta of quantum excitations,
often~called a ``phononic'' limit by analogy with laboratory quantum liquids,
Lorentz symmetry does emerge.
This can be easily shown by virtue of the fluid/gravity analogy~\cite{blv05},
which~was subsequently used to formulate the BEC-spacetime correspondence~\cite{z11appb};
it can also be demonstrated by using dispersion relations~\cite{z20ijmpa,z11pla},
which are generally become deformed in theories with non-exact Lorentz symmetry~\cite{sch17,csa18,ol19,plj20}.

This correspondence states that Lorentz symmetry is approximate,
while
four-dimensional spacetime is an induced phenomenon,
determined by the dynamics of quantum Bose liquid 
moving in Euclidean three-dimensional space.
The latter is only observable  by a certain kind of observer, a~F(ull)-observer. 
Other observers,
R(elativistic)-observers, perceive this superfluid
as a non-removable background, which can be modeled as a four-dimensional pseudo-Riemannian manifold.
What is the difference between these types of observers?

F-observers can perform measurements using objects of
arbitrary momenta and ``see'' the fundamental 
superfluid wavefunction's evolution in  three-dimensional
Euclidean space according to \eq \eqref{e:becgeneq} or an analogue thereof.
On the other hand,
R-observers are restricted
to measuring only small-momentum small-amplitude excitations of the background superfluid.
This is somewhat analogous to listening to acoustic waves (phonons) in the conventional Bose-Einstein condensates,
but~being unaware of higher-energy particles such as photons or~neutrons.  

According to BEC-spacetime correspondence,
a R-observer
``sees'' himself located inside four-dimensional curved spacetime with a pseudo-Riemannian metric.
The latter can be written in Cartesian coordinates as~\cite{z11appb}:  
\be\lb{e:metr1}
g_{\mu\nu} 
\propto
\frac{\dn}{c_s}
\left(
\baa{ccc}
-\left[ c_s^2 - \cf^2 (\vena S)^2 \right] & \vdots & -  \cf \vena~S \\
\cdots\cdots & \cdot & \cdots \\
-  \cf \vena S & \vdots & \textbf{I}
\eaa
\right)
,
\ee
where $\cf = \hbar/m$,
$S = S (\rve,\, t) = - i \ln{\left(\Psi (\rve,\, t) /|\Psi (\rve,\, t)|\right)}$ is a phase of the condensate wavefunction written
in the Madelung representation,
$\Psi = \sqrt\dn \, \exp{(i S)}$,
and
$\textbf{I}$ is a three-dimensional unit matrix.
To~maintain the correct metric signature in \eq \eqref{e:metr1}, 
condition 
$|c_s | > \cf \, |\vena S|$ must be imposed, which indicates 
that $c_s$ is the maximum attainable velocity of test particles 
(i.e.,~small-amplitude excitations of the condensate),
moving along geodesics on this induced spacetime.
Therefore, $c_s$ is the velocity of those excitations of vacuum,
which describe massless particles in the low-momentum limit,
whereas massive test particles move along geodesics of a pseudo-Riemannian manifold with metric \eq \eqref{e:metr1}.
According to a R-observer, they are freely falling, independently of their properties including their rest~mass.

In this approach,
we
interpret Einstein field equations not as differential 
equations for an unknown metric; but as a definition for an induced stress-energy tensor,
describing
some effective matter to which test particles couple.
Therefore,
this would be the gravitating matter observed by a R-observer.
We thus obtain 
\be\lb{e:setdef}
\widetilde{T}_{\mu\nu} 
\equiv
\kappa^{-1}
\left[
R_{\mu\nu} (g) - \frac{1}{2} g_{\mu\nu} R (g)
\right]
,
\ee
where $\kappa = {8 \pi G}/{c_{(0)}^2}$ is the Einstein's gravitational constant.
An example of usage of this procedure will be considered in \Sec~\ref{s:cfs}.
While \eq \eqref{e:setdef} is in fact an assumption,
it should hold not only under the validity of conventional general relativity, but~also in other Lorentz-symmetric theories of gravity which are
linear with respect to the Riemann tensor, 
because the form of Einstein equations is quite universal (up to a conformal transform). 
For other Lorentz-symmetric theories,
whose field equations cannot be transformed into this form, definition \eqref{e:setdef} can be adjusted accordingly.

Furthermore, 
one can see from \eq \eqref{e:cappapp}, that $c_s$ contains 
an unknown function $F (\dn)$.
To~determine its form, let us recall
that one of the relativistic postulates 
implies that velocity $c_s$ 
should not depend on density, at~least
in the classical limit.
More specifically, at~low momenta, this~velocity
should tend to the value $ c_{(0)} \approx c $,
where  
$c = 2.9979 \times 10^{10}$ $\text{cm}\, \text{s}^{-1}$
is 
called the {speed of light in vacuum}, 
for~historical reasons.
Recalling \eq \eqref{e:cappapp}, 
this requirement  can be written
as a differential  equation \citen{z11appb}:
\be\lb{e:Feq}
\dn
|F' (\dn)|
=
m c_s^2
\approx
\text{const} (\dn)
, \ee
where 
$\text{const} (\dn)$ denotes a function which does not depend on density.
The solution of 
this differential equation
is a logarithmic function:
\be\lb{e:oFln}
F (\dn) = - b \ln{(\dn/\dnc)}
, \ee
where  $b$ and $\dnc$ are generally real-valued  functions of coordinates.
The wave Equation \eqref{e:becgeneq}
thus narrows down to
\be\label{e:oF}
i \hbar \partial_t \Psi
=
\left[-\frac{\hbar^2}{2 m} \vena^2
+
V_\text{ext}  (\rve,\, t)
- 
b 
\ln{(|\Psi|^{2}/\dnc)}
\right]\Psi
,
\ee
where $b$ is the nonlinear coupling; $b = b (\rve,\, t)$ in general.
Correspondingly,
{Equation}
~\eqref{e:cappapp} yields
\be\lb{e:cappappln}
p 
= 
- (b/m) \dn
,
\ \
c_s
= 
\sqrt{|b|/m}
,
\ee
thus indicating that logarithmic Bose liquid behaves like barotropic perfect fluid;
but only when one 
neglects quantum corrections, and~assumes classical averaging.
This reaffirms the statement made in the previous section about the place of perfect-fluid
models when it comes to gravitational phenomena.
The way gravity emerges in the superfluid vacuum picture is entirely
different from those models, as~will be demonstrated shortly,
after we have specified our working~model.

Some special cases of \eq \eqref{e:oF}, for~example when $b \to b_0 =$ const,
were extensively studied
in the past, although~not for reasons related to quantum liquids~\cite{ros68,bb76}. 
There were also extensive mathematical studies of these equations, to~mention just some
very recent literature~\cite{ss18,af18,aff19,wz19,bcs19,ct19,wtc19,lzh19,zh20,mm20,aj20,ss20,tb20je,tb20mj}. 

Interestingly, wave equations with logarithmic nonlinearity can be also introduced
into fundamental physics 
independently of relativistic arguments~\cite{z11appb,z10gc,szm16}.
This nonlinearity readily occurs in the theory of open quantum systems, quantum entropy
and information~\cite{yas78,bra91}; as well as in the theory of general condensate-like
materials, for~which characteristic kinetic energies are significantly smaller than interparticle
potentials~\cite{z18zna}.

One example of such a material is helium II, the~superfluid phase of helium-4.
For the latter, the~logarithmic superfluid model is known to have been well verified by experimental data~\cite{z12eb,sz19}.
Among other things, the~logarithmic superfluid model does reproduce the 
sought-after Landau-type spectrum of excitations, discussed in the previous section;
detailed derivations can be found in~\cite{z12eb}. 
One of underlying reasons for such phenomenological success is that the ground-state wavefunction of 
free (trapless) logarithmic liquid
is not a de Broglie plane wave, but~a spatial Gaussian modulated by a de Broglie plane wave.
This explains the liquid's inhomogenization followed by 
the formation of fluid elements or parcels;
which indicates that
such models do describe fluids, rather than gaseous matter~\cite{az11,bo15,z17zna,z18epl,kdk19,z19ijmpb}.

To summarize, a~large number of arguments to date, both theoretical and experimental,
demonstrate the robustness of logarithmic models in the general theory of superfluidity.
In the next \Sec~we shall demonstrate the logarithmic superfluid model's capabilities 
when assuming 
superfluidity of the physical vacuum~itself.

In what follows,
we shall make use of a minimal inhomogeneous model for the logarithmic superfluid
which was proposed in \Ref~\cite{z18zna}, 
based on statistical and thermodynamics arguments.
In the F-observer's picture,
its wave equation can be written as
\be
i\hbar \partial_t \Psi
=
\left[
-\frac{\hbar^{2}}{2 m} \lapl
+V_\text{ext}  (\rve, t)
-
\left(
b_0 - \frac{\tfc}{\rad^2}
\right)\!
 \ln{\!\left(\frac{|\Psi|^{2}}{\dnc} \right)}
\right]\!\Psi
,\label{e:vcm}
\ee
where $\rad  = |\rve| = \sqrt{\rve \cdot \rve}$
is a radius-vector's absolute value,
and $b_0$ and $q$ are real-valued constants.
For definiteness, let us assume that $b_0 > 0$, because~one can always change the overall signs of the nonlinear term
$F (\dn)$ and the corresponding field-theoretical potential $\lanp (\dn)$. 
As always, this wave equation must be supplemented with a normalization condition 
\eqref{e:norm}, boundary and initial conditions of a quantum-mechanical type;
which ensure the fluid interpretation of $\Psi$ \cite{ry99}.

One can show that   
nonlinear coupling $b = b (\rve) = b_0 - \tfc/\rad^2$
is a linear function of the quantum temperature $\tps$, 
which is defined as 
a thermodynamic conjugate
of quantum information entropy sometimes dubbed as the Everett-Hirschman information entropy.
The latter can be written as
$ 
S_\Psi = 
-  
\langle \Psi | \ln{(|\Psi|^2/\dnc)}|\Psi \rangle
=
-\int_\vol |\Psi|^2 \ln{(|\Psi|^2/\dnc)} \, \dvol
$, 
where a factor $1/\dnc$ is introduced for the sake of correct dimensionality,
and can be absorbed into an additive constant due to the normalization condition \eqref{e:norm}.
Therefore, one can expect that the thermodynamical parameters 
\be\lb{e:bqtem}
b_0 = b_0 (\tps), \ \tfc = \tfc (\tps)
\ee
are constant at a fixed temperature $\tps$.
Thus, for~a trapless version of the model \eqref{e:vcm}
we have four parameters, but~only two of them, $m$ and $\dnc$,
are  \textit{a priori} fixed, whereas the other two, $b_0$ and $\tfc$,
can vary depending on  
the~environment.


\scn{Induced Gravitational Potential}{s:gr}
Invoking model \eqref{e:vcm}, while neglecting quantum fluctuations, let us assume that physical vacuum is a collective
quantum state described
by wavefunction $\Psi = \wfv (\rve,\, t)$,
which forms a self-gravitating configuration with a center at $\rve = 0$.
Therefore, for~this state, the~solution of \eq \eqref{e:vcm} is equivalent to the solution
of the linear Schr\"odinger equation,
\be
i\hbar \partial_t \Psi
=
\left[
-\frac{\hbar^{2}}{2 m} \lapl
+V_\text{eff} (\rve,\, t)
\right]
\Psi
,\label{e:vcmef}
\ee
for a particle of mass $m$
driven by an effective potential 
\ba
V_\text{eff} 
(\rve, t)
&=&
V_\text{ext}  (\rve, t)
-
b 
\ln{\!\left(\frac{|\wfv (\rve, t)
|^{2}}{\dnc} \right)}
\nn\\
&=&
V_\text{ext}  (\rve, t)
-
\left(
b_0 - \frac{\tfc}{\rad^2}
\right)\!
 \ln{\!\left(\frac{|\wfv (\rve, t)
|^{2}}{\dnc} \right)}\!,~~~~~
\lb{e:vind}
\ea
when written in Cartesian coordinates~\cite{z18zna}.
If working in curvilinear coordinates, the~last formula must be supplemented with terms
which arise after separating out the angular variables in the wave~equation.

In the absence of quantum excitations and other interactions, it is natural to associate 
this effective quantum-mechanical potential 
with the only non-removable fundamental interaction that we know of: gravity.
This interpretation will be further justified in \Sec~\ref{s:phy}.
Therefore, in~Cartesian coordinates one can write the induced gravitational potential as
\be\lb{e:igrav}
\Phi (\rve, t)
=
- \frac{1}{m} 
V_\text{eff} (\rve, t) 
=
\frac{1}{m} 
\left(
b_0 - \frac{\tfc}{\rad^2}
\right)
\ln{\!\left(\frac{|\wfv (\rve, t)|^{2}}{\dnc} \right)}
,
\ee
where we assume that the background superfluid is trapless, i.e.,~we set $V_\text{ext} = 0$.
It should also be  remembered that in curvilinear coordinates, 
this formula must be modified according to the remark after \eq \eqref{e:vind}; but for now we 
shall disregard any anisotropy
and~rotation.

It should be noticed that if one regards this potential as a multiplication operator 
then its quantum-mechanical average would be related to
the~Everett-Hirschman information entropy discussed
in the previous section:
$
\langle
\Phi 
\rangle
\sim \tps\left\langle \Psi\right|\, \ln{(|\Psi|^{2})} \left|\Psi \right\rangle
\sim \tps
S_\Psi
$. 
This not only makes theories of entropic gravity
(which are essentially based on the ideas of Bekenstein, Hawking, Jacobson and others) a subset of the logarithmic superfluid vacuum 
approach, but~also endows them with an underlying physical meaning and origin of the entropy~implied.

We can see that the induced potential maintains its form as long as the physical vacuum stays in the state 
$|\wfv \rangle$.
If the vacuum were to transition into a different state, then it would change its wavefunction;
hence  the induced gravitational potential 
would also change.
We expect that our vacuum is currently in a stable state,
which is close to a ground state 
or at least to a metastable state, with~a sufficiently large lifetime.  
It is thus natural to assume that the state 
$|\wfv \rangle$
is stationary and rotationally~invariant.

As we established earlier, the~wavefunction describing such a state
should be the solution
of a quantum wave equation containing logarithmic nonlinearity.
In the case of trivial spatial topology and infinite extent,
the amplitude
of such a solution is known to be the product of a Gaussian function,
which was mentioned in the previous section,
and a conventional quantum-mechanical part, which~is a product of an exponential function,
power function and a polynomial.
Thus we can write the amplitude's general form as:
\be\lb{e:avex}
|\wfv | =
\sqrt{\dnc}\,
\left(
\frac{\rad}{\lnc}
\right)^{\pwr_0/2}
P (\rad)
\,
\exp{\!
\left(
-
\frac{a_2 }{2 \lnc^2} \rad^2
+
\frac{a_1}{2 \lnc} \rad
+ 
\frac{a_0}{2}
\right)
}
,
\ee
where 
$P (\rad)$ is a polynomial function,
$\pwr_0$
and $a$'s are constants,
and $\lnc = (m/\dnc)^{1/3}$
is a classical characteristic length scale of the logarithmic nonlinearity
(alternatively, one can choose $\lnc$ being equal to the quantum characteristic length,
$\hbar/\sqrt{m b_0}$, which might be more useful for $\hbar$-expansion techniques).
If quantum liquid occupies an infinite spatial domain then the 
normalization condition \eqref{e:norm}  
requires
\be
a_2 > 0
,
\ee
which is also confirmed by analytical and numerical studies of differential equations with 
logarithmic nonlinearity of 
various types~\cite{bb76,ss18,af18,bcs19,lzh19,z18zna,ss20}.

Both the form of a function $P (\rad)$ and the values of $\pwr_0$ and $a$'s 
must be determined by a solution of an eigenvalue problem
for the
wave equation 
under normalization 
and boundary conditions.
At~this stage, those conditions are not yet precisely known;
even if they were, we do not yet know which quantum state our vacuum is currently in.
Therefore, these constants' values remain theoretically unknown at this stage,
yet  can be determined empirically. 

Furthermore, for~the sake of simplicity, 
let us approximate
the power-polynomial term $(\rad/\lnc)^{\pwr_0/2} P (\rad)$, 
by the single power function $(\rad/\lnc)^{\pwr/2}$, where the constant $\pwr$ 
is 
the best fitting parameter.
Therefore, we can approximately rewrite \eq \eqref{e:avex} as
\be\lb{e:ava}
|\wfv |^2 \approx
\dnc\, 
\exp{\!
\left[
-
\frac{a_2 }{\lnc^2} \rad^2
+
\frac{a_1 }{\lnc} \rad
+ \pwr 
\ln{\!\left(
\frac{\rad}{\lnc}
\right)}
+ a_0
\right]
}
,
\ee
which is more convenient for further analytical studies than the original expression \eqref{e:avex}.
From the empirical point of view, the~function \eqref{e:ava} can be considered as a
trial function, whose parameters can be fixed using experimental data
following the procedure we  describe~below. 

For the trial solution \eqref{e:ava},
the
normalization condition \eqref{e:norm} immediately imposes a constraint for one 
of its parameters:
\be
\text{exp}{(a_0)}
\approx
\frac{\Mass
a_2^{(\pwr +3)/2}
}{2 \pi m}
\left[
\Auxfun{\frac{3}{2}}{\frac{1}{2}}
+
\frac{a_1}{\sqrt{a_2}}
\Auxfun{2}{\frac{3}{2}}
\right]^{-1}
,
\ee
where we introduced an auxiliary function
$
\Auxfun{a}{b}=
\Gamma\left(a + \pwr/2 \right)
\Kummer{a+ \pwr/2}{b}{a_1^2/4 a_2}
$,
where
$\Gamma (a)$ and $\Kummer{a}{b}{z}$ are the gamma function and Kummer confluent hypergeometric function,
respectively.
If values of $a$'s and $\pwr$ are determined, e.g.,~empirically, then this formula
can be used to estimate the ratio $\Mass/m$.

Furthermore,
by
substituting the trial solution \eqref{e:ava} into the definition \eqref{e:igrav},
we derive the induced gravitational potential as a sum of seven terms:
\ba
\grpott (\rad)
&=&
\grsmi
(\rad)
+
\grmic
(\rad)
+
\grnew
(\rad)
\nn\\&&
+
\grgal
(\rad)
+
\grmgal
(\rad)
+
\grcos
(\rad)
+
\grpott_0 
,\lb{e:grev}
\ea
where
\ba
\grsmi (\rad) 
&=&
-\frac{\pwr \, q}{m}
\frac{\ln{(\rad/\lnc)}}{\rad^2}
=
- \zeta_{\pwr q} c_b^2
\frac{\lsmi^2 \ln{(\rad/\lnc)}}{\rad^2}, 
~~~\lb{e:grsmi}\\
\grmic (\rad) 
&=&
-\frac{a_0 \, q}{m}
\frac{1}{\rad^2}
=
- \zeta_{a_0 q} c_b^2
\frac{\lmic^2}{\rad^2}
\, , \lb{e:grmic}\\
\grnew (\rad) 
&=&
-\frac{a_1 q}{m \lnc} \frac{1}{\rad} =
- \frac{G M}{\rad}
, \lb{e:grnew}\\
\grgal (\rad) 
&=&
\frac{\pwr \, b_0}{m}
\ln{(\rad/\lnc)}
=
c_b^2\,
\pwr \ln{(\rad/\lnc)}
,  \lb{e:grgal}\\
\grmgal (\rad) 
&=&
\frac{a_1 b_0}{m \lnc}
\rad
=
\zeta_{a_1} c_b^2
\frac{\rad}{\lmgl}
,  \lb{e:grmgl}\\
\grcos (\rad) 
&=&
-\frac{a_2 b_0}{m \lnc^2}
\rad^2
=
- c_b^2
\frac{\rad^2}{\lcos^2}
,    \lb{e:grcos}
\ea
and
\be
\grpott_0 
=
\frac{1}{m} 
\left(
a_0 b_0 
+ \frac{a_2 \tfc}{\lnc^2}
\right)
=
\frac{1}{m} 
\left(
a_0 b_0 
+ \frac{\tfc}{\lcos^2}
\right)
\,  \lb{e:gr0}
\ee
is the additive constant.
Here, and~throughout the paper, we denote the sign functions
by $\zeta$'s: 
$\zeta_\alpha = \Sign{\alpha}$,
and use the following notations:
\ba
&&
c_b = \sqrt{\frac{b_0}{m}}, \
G M 
= \frac{a_1 q}{m \lnc}, 
\
\lsmi = \sqrt{\frac{|\pwr\, q|}{b_0}}
, 
\nn\\&& 
\lmic = \sqrt{\frac{|a_0 q|}{b_0}}
,\
\lmgl = \frac{\lnc}{|a_1|}
= \frac{|q|}{m G M}
,
\lb{e:lengr}\\&&  
\lcos =  \frac{\lnc}{\sqrt{a_2}}
, \
\lpwr = \frac{\pwr \lnc}{a_1}
= \frac{\pwr q}{m G M}
,
\nn
\ea
where
$G$ is the Newton's gravitational constant as per~usual.

Furthermore, 
Lorentz symmetry emerges
in the ``phononic'' low-momentum limit of the theory, as discussed in the previous section.
Therefore,
a R-observer 
would perceive the gravity induced by potential \eqref{e:grev}
as curved four-dimensional spacetime,
which is a local perturbation (not necessarily small)
of the background flow metric,
such as the one derived in \Sec~5.3 of \Ref~\cite{z11appb},
see \Sec~\ref{s:cfs} below.
In a rotationally invariant case,
the line element of this spacetime
can be written in the Newtonian gauge;
if $\grpott (r)/c_{(0)}^2 \ll 1$, 
then it can be approximately 
rewritten in the form
\be\lb{e:metrst}
d s^2 
\approx
- c_{(0)}^2\!\left[
1 + 
\frac{2 \grpott (r)}{c_{(0)}^2}
\right] d t^2
+ \frac{d r^2}{1 + 2 \grpott  (r)/c_{(0)}^2} + 
R^2 (r) d \mesph
,
\ee
where 
$R (r) = r \left[1 + {\cal O} \left(\grpott (r)/c_{(0)}^2\right)\right] \approx r$,
$d\, \mesph = d\theta^2 + \Sin{2}{\theta} \, d \varphi^2$
is the line element of a unit two-sphere,
and a leading-order approximation
with respect
to the Planck constant is implied, 
as~usual.
The mapping \eqref{e:metrst} is valid for regions where 
the induced metric maintains a signature `$- + + +$',
and its matrix is non-singular.
In other regions, such as close vicinities of spacetime singularities
or horizons,
the relativistic approximation is likely to fall
outside its applicability range,
thus it should be replaced with
the F-observer's description of~reality.

The main 
simplifying assumptions and approximations underlying
the derivation of our gravitational potential
are summarized and enumerated in the Appendix \ref{s:aa}.

\scn{Physical Interpretation}{s:phy}
It should be noticed that if we did not have a logarithm
in the original model \eqref{e:vcm}, then~in \mbox{\eqs \eqref{e:vind}
and \eqref{e:igrav}},
then
we would not have arrived at the polynomial functions
in \mbox{\eqs~\eqref{e:grev}--\eqref{e:grcos}},
which are easily recognizable.
This reaffirms our expectations that the underlying model
can be successfully confirmed by experiment;
but first those functions must be endowed with precise physical~meaning. 

In this \Sec,
we shall assign a physical interpretation to each term of the derived gravitational potential.
For the sake of brevity,
we shall be omitting an additive constant $\grpott_0$,
assuming that it is small compared to $c_{(0)}^2$.

\sscn{Potential $\grnew$ and gravitational mass generation}{s:pnew}

We begin with term \eqref{e:grnew}, which has the most obvious meaning.
In a non-relativistic picture, it~represents Newton's model of gravity.
According
to the BEC-spacetime correspondence manifested through
the mapping \eqref{e:metrst},
a R-observer can observe an effect of the potential $\grnew$ by
measuring probe particles moving along geodesics in the Schwarzschild spacetime: 
\be\lb{e:meschw}
d s^2_{(\text{N})}
\approx
- c_{(0)}^2 
\left(
1 - \frac{r_H}{r}
\right) d t^2
+ \frac{d r^2}{1 - r_H / r} + 
r^2 d \mesph
,
\ee
where
$r_H = 2 G M/c_{(0)}^2$ is the Schwarzschild radius.
Therefore,
in absence of asymptotically non-vanishing terms, $M$ can be interpreted as the gravitational mass of
the~configuration.
 
This mass can be expressed in terms of superfluid parameters as
\be\lb{e:schrad}
r_H =
\frac{2 a_1 q}{m c_{(0)}^2 \lnc}
,
\ \
\Sign{a_1 q} = 
\left\{
\baa{rll}
1 & &\text{gravity},\\
-1 & &\text{anti-gravity},
\eaa
\right.
\ee
thus assigning physical meaning to a combination of parameters $a_1 q/m \lnc$.
In particular, 
one can see
that
a sign of the product $a_1 q$ 
determines
whether the $\grnew$ interaction is attractive (gravity) or repulsive (anti-gravity).
For most systems that we know of, anti-gravitational effects have not yet been  observed, 
therefore one can assume that $M > 0$ or 
\be
a_1 q > 0
\ee 
from now~on. 

Nevertheless, 
it should be emphasized that the anti-gravity case is not 
\textit{a priori}
forbidden in superfluid vacuum theory.
Indeed, the~spacetime singularity occurs at 
$r = 0$ in a relativistic picture only, which poses certain issues for a R-observer,
especially in the case of anti-gravity when a singularity is not covered by an event horizon
(the existence of naked singularities is often doubted, on~grounds of the cosmic censorship hypothesis).
However, a~F-observer would see no singular behavior in either case, because~the wavefunction 
$\wfv$
remains
regular and normalizable at each point of space and at any given time -- as it should be in a quantum-mechanical theory. 
This reaffirms the fact that spacetime singularities are an artifact of incomplete information accessible
to observers operating with 
relativistic particles~\cite{z11appb}.

Thus, the~mapping from \eqs \eqref{e:grnew}--\eqref{e:meschw}
can be used to reformulate black hole phenomena in the language of continuum mechanics and 
the theory of superfluidity;
which can resolve certain long-standing problems occurring in the relativistic theory of gravity.
For instance,
neglecting asymptotically non-flat terms for simplicity,
one can view \eqs \eqref{e:grnew}, \eqref{e:meschw} and \eqref{e:schrad} 
as the gravitational mass generation mechanism:
such mass is not a fundamental notion,
but a composite quantum phenomenon induced by the background superfluid's dynamics 
(through the elementary inertial mass $m$ and
critical density $\dnc$), its quantum temperature (through $q$),
and an
exponential part of the condensate's wavefunction (through $a_1$).
Such a mechanism can be thus considered as the quantum-mechanical version of the Mach principle~\cite{z11appb}.
For example,
if either $a_1$ or $q$ vanish, then the system would not possess any gravitational mass,
but it still would be gravitating in a non-Newtonian way, if~other potentials from \eq \eqref{e:grev} are non-zero.

\sscn{Potential $\grmic$ and abelian charges}{s:pmic}
Equation \eqref{e:grmic} represents another potential which can be easily recognized.
According
to the mapping \eqref{e:metrst},
potential $\grmic$ is observed by a R-observer as Reissner-Nordstr\"om spacetime,
when~taken together with the $\grnew$ potential:
\ba
d s^2_{(\text{N}+\text{RN})}
&\approx&
- c_{(0)}^2 
\left(
1 - \frac{r_H}{r} + \frac{r_Q^2}{r^2}
\right) d t^2
\nn\\&&
+ \frac{d r^2}{1 - r_H / r + r_Q^2/r^2} 
+ 
r^2 d \mesph
,\lb{e:memic}
\ea
where $r_Q$
is a characteristic length scale:
\be
r_Q
=
\sqrt 2 \lmic 
\frac{ c_b  }{c_{(0)}}
= 
\sqrt{\frac{2 |a_0 q|}{m c_{(0)}^2}}
,
\ee 
provided \
$a_0 q < 0$.

The Reissner-Nordstr\"om metric is known to be associated with the gravitational field caused
by a charge related to an abelian group. 
An example would be an electric charge $Q$,
which is related to
the abelian group $U(1)$ of electromagnetism.
This charge can be thus revealed through the formula
\be
Q^2 = \frac{r_Q^2 c_{(0)}^4}{k_e G}
\approx
\frac{2 | a_0 q| c^2}{k_e G m}
,
\ee
where $k_e$ is the Coulomb constant.
In other words, 
potential $\grmic$ describes, together with $\grnew$, 
the~gravitational field created by an object of a charge $Q$ and gravitational mass $M$.

Thus, from~a F-observer's viewpoint,
an electrical charge is not
an elementary notion,
but a composite quantum phenomenon, induced by the background superfluid's dynamics (through the elementary inertial mass $m$), its quantum temperature (through $q$),
and 
overall constant coefficient of the condensate's wavefunction (through $a_0$).

It should be noted also that $\grmic$ is a short-range potential,
therefore, it becomes substantial only at those microscopical length scales, 
of an order
$r_{QH} = r_Q^2/r_H = \lnc\,|a_0/a_1|$ or below.
Since  $r_Q < r_H$ for most objects we know of,
we have $0 \leqslant r_{QH} < r_H$.
Thus, those scales would be causally inaccessible to a R-observer, but~a F-observer would have no problem accessing them, per~usual.

\sscn{Potential $\grsmi$ and strong gravity}{s:psmi}
As the distance from a gravitating center decreases,
it is  term \eqref{e:grsmi} which eventually predominates. 
According
to the mapping \eqref{e:metrst},
this potential $\grsmi$,
when taken together with the $\grnew$ and $\grmic$ potentials,
induces spacetime with the line element:
\bw
\ba
d s^2_{(\text{N}+\text{RN}+\text{smi})}
\!&\approx&\!
- c_{(0)}^2\!\left[
1 - \frac{r_H}{r} + \frac{r_Q^2}{r^2}
-
\zeta_{\pwr q}
\frac{\rad_W^2}{\rad^2}
\ln{\left(\frac{\rad}{\lnc}\right)}
\right]\! d t^2
+ \frac{d r^2}{1 - r_H / r + r_Q^2/r^2
-
\zeta_{\pwr q}
\rad_W^2
\ln{\left(\rad/\lnc\right)}/\rad^2
} 
+ 
r^2 d \mesph,~~~~
\lb{e:mesmi}
\ea
\ew
where $r_W$
is a characteristic length scale:
\be
r_W
=
\sqrt 2
\lsmi
\frac{ c_b  }{c_{(0)}}
= 
\sqrt{
\frac{2 |\pwr\, q|}{m c_{(0)}^2}
}
.
\ee 

The potential  $\grsmi$ has a distinctive property:
unlike other sub-Newtonian potentials in \mbox{\eq~\eqref{e:grev}}, 
it can switch between repulsive and attractive regimes,
depending on whether the distance is larger or smaller than $\lnc$.

The
magnitudes of $\grsmi$ and $\grmic$ become comparable
at two values of $r$,
shown by the formula
\be
r_{WQ}^{(\pm)} =
\lnc 
\exp{(\pm r_Q^2/r_W^2)}
=
\lnc 
\exp{(\pm |a_0/\pwr|)}
,
\ee
which indicates
that $|\grsmi |$ overtakes $|\grmic |$ 
either at  $r < r_{WQ}^{(-)}$ or at $r > r_{WQ}^{(+)}$.
If $|a_0/\pwr|$ is large then $r_{WQ}^{(+)}$ is exponentially large and
$r_{WQ}^{(-)}$ is exponentially~small. 

Magnitudes of $\grsmi$ and $\grnew$ become comparable
at a certain value of $r$,
which is
shown by the~formula
\ba
r_{WM} 
&=& 
-
\ell_W
\Prlog{
-
\lnc/\ell_W
}
\leqslant \ell_W,
\ea
where
$
\ell_W 
\equiv 
{r_W^2}/{|r_H|}
=
|\lpwr|
$,
and by $\Prlog a$ we denoted the Lambert function.
The function $|\grsmi (\rad)|$ becomes larger than $|\grnew (\rad)|$ 
at  $r < r_{WM} \leqslant |\lpwr| $.

Furthermore,  
considering $\grsmi$ as a perturbation of $\grnew$,
one can deduce that 
the effective gravitational coupling,
\be\lb{e:Geff}
G_\text{eff} \approx 
G
\left[
1 + 
\zeta_{\pwr q} 
\lpwr
\frac{\ln{(r/\lnc)}}{r} 
\right] 
, 
\ee 
becomes larger as $r$ gets smaller.
Here
an approximation symbol reminds us that we are working with 
wavefunction \eqref{e:ava}, and~assume a leading-order approximation
with respect
to the Planck constant.

One can see that 
gravity naturally becomes stronger at shorter scales, 
without introducing any additional effects or matter,
which suggests the way towards resolving the hierarchy problem. 
Indeed, the~latter states the large discrepancy between magnitudes of forces of Standard Model 
interactions and classical gravity,
but 
in this case, 
the gravitational force 
grows stronger than inverse square as $r$ decreases, as~$\ln r/r^3$.
Moreover, 
it is possible to have even stronger short-length behaviour in our approach:
if one goes beyond a minimal model \eqref{e:vcm} and generalizes  
nonlinear coupling to a series: $b (\rve) = b_0 - \tfc/\rad^2 + q_1/\rad^3 + ...$,
then this will induce terms $\ln r/r^4$,  $\ln r/r^5$, \textit{et cetera}.

On the other hand, as~$r$ grows, the~function 
$G_\text{eff} $
converges to $G$,
especially if $|\lpwr |$ or $1/\lnc$ are sufficiently small;
which makes $G_\text{eff}$ approximately constant for a large range of values of $r$.

\sscn{Potential $\grcos$ and de Sitter spacetime}{s:pcos}
Let us turn our attention to terms
which do not tend to zero at $\rad \to \infty$.
The physical meaning of one of these 
terms, given by \eq \eqref{e:grcos},
becomes clear upon
using the mapping \eqref{e:metrst}: 
\be\lb{e:meds}
d s^2_{(\text{dS})}
\approx
- c_{(0)}^2 
\left(
1 - \frac{r^2}{\rcos^2}
\right) d t^2
+ \frac{d r^2}{1 - r^2 / \rcos^2} + 
r^2 d \mesph
,
\ee
where 
\be
\rcos = 
\frac{1}{\sqrt 2}
\frac{c_{(0)}}{c_b}
\lcos
= 
 \lnc
\sqrt{\frac{m  c_{(0)}^2}{2 a_2 b_0}}
\ee 
is a radius of de Sitter~horizon.

This metric represents de Sitter spacetime (written
in static coordinates),
which belongs to a class of  Friedmann-Lema\^itre-Robertson-Walker (FLRW)
spacetimes.
Indeed, by~applying a coordinate transformation
$\tau - t = \tau_\text{dS} 
\ln{\!\left(1 - r^2/\rcos^2
\right)}
$,
$
\varrho = \alpha_0^{-1} r
\exp{\!\left(
- t/2 \tau_\text{dS}
\right)}
\left(1 - r^2/\rcos^2 
\right)^{-1/2}
$,
one can rewrite \eq \eqref{e:meds} in isotropic coordinates
\be\lb{e:meds2}
d s^2_{(\text{dS})}
\to
- c_{(0)}^2 d \tau^2
+ 
\alpha_0^2 
\exp{\!\left(
\frac{\tau}{\tau_\text{dS}}
\right)
}
\left(
d \varrho^2 + \varrho^2 d \mesph
\right)
,
\ee
where
$\tau_\text{dS} = \rcos/ 2 c_{(0)}$
and $\alpha_0$ is an integration~constant.

The physical implications of the term $\grcos$
will be further examined in \Sec~\ref{s:uexp}.

\sscn{Potentials $\grmgal$ and $\grgal$ and gravity on astronomical scales}{s:pmgal}
The remaining asymptotically non-vanishing potentials are given by \eqs \eqref{e:grgal} and \eqref{e:grmgl}. 
With respect to the dependence upon radial distance from the gravitating center, 
they occupy an intermediate
place between de Sitter term \eqref{e:grcos} and Newtonian potential \eqref{e:grnew}.
It is thus natural to expect that
these terms are responsible for large scale dynamics --
from galaxies (a kiloparsec scale) to metagalactic objects, such as voids and superclusters
(a megaparsec scale).

According to the mapping \eqref{e:metrst},
the terms $\grgal$ and $\grmgal$ modify  de Sitter metric \eqref{e:meds}:
causing the resulting spacetime to be asymptotically de Sitter only.
The line element, which corresponds to the terms $\grcos$,
$\grmgal$ and $\grgal$ taken together, 
can be written in static coordinates as
\bw
\ba
d s^2_{(\text{gal+mgl+dS})}
&\approx&
- c_{(0)}^2 
\left[
1 
+ \beta_\pwr^2 
\ln{\!\left(
\frac{\rad}{\lnc}
\right)} 
+ \frac{r}{R_\ssmgl} - \frac{r^2}{\rcos^2}
\right] d t^2
+ \frac{d r^2}{
1 
+ 
\beta_\pwr^2 \ln{\left(\rad/\lnc\right)}+ r/R_\ssmgl - r^2 / \rcos^2
} + 
r^2 d \mesph
,
\lb{e:medsa}
\ea 
\ew
where 
$ 
\beta_\pwr
=
\sqrt{
{2 \pwr \, b_0}/{m c_{(0)}^2}
}
=
\sqrt{
2 \pwr
}
{c_b}/{c_{(0)}}
$, 
and
\be
R_\ssmgl
=
\frac{m c_{(0)}^2 \lnc}{2 a_1 b_0} 
=
\frac{\zeta_{a_1}}{2}\frac{c_{(0)}^2}{c_b^2}
\lmgl
\ee 
is a characteristic length scale constant,
which can be positive or negative depending on the sign of $a_1$.
Note that in \eq \eqref{e:medsa},
we included the $\grcos$-induced (de Sitter) term 
to remind us that at large $r$ we still have a spacetime of a
FLRW~type.

Interestingly, linear terms in metrics 
occur also in an alternative theory of gravity, Weyl gravity~\mbox{\cite{mk89,man93}}.
This coincidence can be explained by conformal symmetry, which often emerges in logarithmic models
at the relativistic limit,
see for example \Sec~\ref{s:cfs}. 
However, Weyl gravity does not produce a logarithmic term in metric,
while in our theory it is induced by $\grgal$.
This term is responsible for the flat rotation curves phenomenon in galaxies,
which will be discussed, along with other astronomical implications of the terms
$\grmgal$ and $\grgal$, in~further detail in \Sec~\ref{s:frc}.

It is also useful to know that the  metric induced by the term \eqref{e:grmgl} alone,
\be
d s^2_{(\text{mgl})}
\approx
- c_{(0)}^2 
\left(
1 
+ \frac{r}{R_\ssmgl} 
\right) d t^2
+ \frac{d r^2}{
1 
+ r/R_\ssmgl 
} + 
r^2 d \mesph
,
\lb{e:memgl}
\ee 
transforms to the conformally FLRW-type metric
\bw
\ba
d s^2_{(\text{mgl})}
&\to&
\frac{
(1+\varrho/4 R_\ssmgl)^2
}{
a_\ssmgl^2 (\tau)\, (1-\varrho/4 R_\ssmgl)^2
}
\left[
- c_{(0)}^2  d \tau^2
+ \frac{
a_\ssmgl^2 (\tau)
}{(1-\varrho^2/16 R_\ssmgl^2)^2
}
\left(
d \varrho^2 + \varrho^2 d \mesph
\right)
\right]
,
\lb{e:memgl2}
\ea 
\ew
upon applying the coordinate transformation
$r = \varrho/(1-\varrho/4 R_\ssmgl)^2$,
$t = \int d \tau/a_\ssmgl (\tau)$.
For R-observers,
this represents a surrounding
homogeneous and
isotropic spacetime with 
scale factor $a_\ssmgl (\tau)$
and negative-definite spatial scalar curvature,
$- 1/(2 R_\ssmgl)^2$.

Finally, one could mention that
potentials of type \eqref{e:grgal} were studied, albeit in the context
of a linear Schr\"odinger equation,
in \Refs~\cite{zs18a,zs18b}.

\scn{Density of Effective Gravitating Matter 
}{s:den}
In this \Sec, let us derive spherically-symmetric density profiles
of effective gravitating matter $\dnind$, which 
formally corresponds to the superfluid-vacuum induced potential \eqref{e:grev}.
Contrary to the directly observable value of orbital velocity derived in \Sec~\ref{s:frc},
such density 
depends more substantially
on the choice of an observer.
It can be 
defined
either via the Lorentz-covariant definition
~\eqref{e:setdef},
or via
the Poisson equation,
which is a non-relativistic version of \eq \eqref{e:setdef}.
Correspondingly, 
one would obtain different results,
which will be discussed~below. 

In accordance with the last paragraph of Appendix 
\ref{s:aa},
this \Sec's computations cannot take into account any secondary induced matter,
such as the equilibrium configurations of mass-energy emerging as a result of interaction between scalar and tensor modes
of superfluid vacuum's excitations.
One can show that such equilibria do exist, manifesting themselves in a form of general relativistic nonsingular horizon-free stellar-like objects
or particle-like Q-balls, 
therefore,
in reality such objects would definitely contribute to the density associated with dark matter.
In other words, here
we are restricting ourselves to background values of~density.

In this \Sec~only,
we shall be temporarily assuming
that
$ 
M = M (\rad)
$, 
otherwise the corresponding contribution to the density profile would be identically zero, 
and thus~non-indicative.

\sscn{Galilean symmetry 
}{s:denn}
In this case,
one defines an effective matter density by virtue of the Poisson equation.
Taking the whole potential \eqref{e:grev} and assuming spherical symmetry, 
we obtain
\bw 
\ba
\dnind
(\rad)
&\equiv&
\frac{1}{4 \pi G}
\left[
\grpott'' (\rad)
+
\frac{2}{\rad} \grpott' (\rad)
\right]
= 
\rho_{\text{smi}} (\rad)
+
\rho_{\text{RN}} (\rad)
+
\rho_{\text{N}} (\rad)
+
\rho_{\text{gal}} (\rad)
+
\rho_{\text{mgl}} (\rad)
+
\rhcos (\rad)
,
\lb{e:rhodm}
\ea
\ew
where
\ba
\rho_{\text{smi}}
(\rad)
&=&
\frac{3 \pwr \, q}{4 \pi G m r^4}
\left[1 -
\frac{2}{3}
\ln{\left(
\frac{\rad}{\lnc}
\right)}
\right]
,\lb{e:rhosmi}\\
\rho_{\text{RN}}
(\rad)
&=&
-
\frac{a_0 q}{2 \pi G m r^4}
=
\frac{k_e Q^2}{4 \pi r^4}
,\lb{e:rhomic}\\
\rho_{\text{N}}
(\rad)
&=&
-
\frac{1}{4 \pi r}
M''(r)
,\lb{e:rhonew}\\
\rho_{\text{gal}}
(\rad)
&=&
\frac{\pwr \, b_0}{4 \pi G m r^2}
=
\frac{v_\text{gal}^2}{4 \pi G r^2}
,\lb{e:rhogal}\\
\rho_{\text{mgl}}
(\rad)
&=&
\frac{a_1 \, b_0}{2 \pi G m \lnc r}
=
\frac{\pwr \, b_0}{2 \pi G m \lpwr r}
,\lb{e:rhomgl}\\
\rho_{\text{dS}}
(\rad)
&=&
-
\frac{3 a_2 \, b_0}{2 \pi G m \lnc^2}
=
-
\frac{3 b_0}{2 \pi G m \lcos^2}
= \text{const}
,\lb{e:rhocos}
\ea
whereby 
the 
sum 
of the last three densities
can be regarded as a density corresponding to the astronomical-scale ``dark matter'' and ``dark energy'',
which will be further justified in \mbox{\Secs \ref{s:frc} and \ref{s:uexp}}.

Furthermore,
these
formulae were derived on the assumption
that the gravitational coupling constant $G$ is the same for all length scales;
which is valid when any influence from the term \eqref{e:grsmi} can be disregarded.
However, this term might cause an additional effect, discussed in \Sec~\ref{s:psmi}:
it makes the gravitational coupling constant vary with distance.
If this does happen, then in \eqs \eqref{e:rhodm}--\eqref{e:rhocos}  
one should replace $G$  with a running constant given by \eq \eqref{e:Geff}.

\sscn{Lorentz symmetry 
}{s:denr}
In this case,
one defines effective matter density by virtue of  Einstein field
equations with the induced stress-energy tensor defined by \eq \eqref{e:setdef}.
Using it together with \eq \eqref{e:grev} and metric~\eqref{e:metrst},
we obtain
\bw 
\ba
\dnindr
(\rad)
&\equiv&
-
\frac{1}{4 \pi G r}
\left[
\grpott' (\rad)
+
\frac{1}{\rad} \grpott (\rad)
\right]
= 
\dnir_{\text{smi}} (\rad)
+
\dnir_{\text{RN}} (\rad)
+
\dnir_{\text{N}} (\rad)
+
\dnir_{\text{gal}} (\rad)
+
\dnir_{\text{mgl}} (\rad)
+
\dnir_{\text{dS}} (\rad)
+
\dnir_{\grpott_0} (\rad)
,
\lb{e:rhodmr}
\ea
\ew
where
\ba
\dnir_{\text{smi}}
(\rad)
&=&
\frac{\pwr \, q}{4 \pi G m r^4}
\left[1 -
\ln{\left(
\frac{\rad}{\lnc}
\right)}
\right]
,\lb{e:rhosmir}\\
\dnir_{\text{RN}}
(\rad)
&=&
-
\frac{a_0 q}{4 \pi G m r^4}
=
\frac{k_e Q^2}{8 \pi r^4}
,\lb{e:rhomicr}\\
\dnir_{\text{N}}
(\rad)
&=&
\frac{1}{4 \pi r^2}
M'(r)
,\lb{e:rhonewr}\\
\dnir_{\text{gal}}
(\rad)
&=&
-
\frac{\pwr \, b_0}{4 \pi G m r^2}
\left[1 +
\ln{\left(
\frac{\rad}{\lnc}
\right)}
\right]
,\lb{e:rhogalr}\\
\dnir_{\text{mgl}}
(\rad)
&=&
-
\frac{a_1 \, b_0}{2 \pi G m \lnc r}
=
-
\frac{\pwr \, b_0}{2 \pi G m \lpwr r}
,\lb{e:rhomglr}\\
\dnir_{\text{dS}}
(\rad)
&=&
\frac{3 a_2 \, b_0}{4 \pi G m \lnc^2}
=
\frac{3  b_0}{4 \pi G m \lcos^2}
= \text{const}
,\lb{e:rhocosr}\\
\dnir_{\grpott_0}
(\rad)
&=&
-
\frac{\grpot_0}{4 \pi G r^2} 
=
-
\frac{1}{4 \pi G m r^2} 
\left(
a_0 b_0 
+ \frac{\tfc}{\lcos^2}
\right)\!,~~~
\lb{e:rhophior}
\ea
whereby 
the 
sum 
of last four densities
can be regarded corresponding to the astronomical-scale ``dark matter'' and ``dark energy'',
which will be further justified in \Secs \ref{s:frc} and \ref{s:uexp}.
Density \eqref{e:rhophior}
is a somewhat surprising contribution, because~it corresponds to the constant term $\grpot_0$ in \eq \eqref{e:grev},
which is not supposed to affect trajectories; at least, in~classical mechanics.
Unless its presence can be confirmed by observations,
it must be regarded as a gauge term, or~as an artifact of approximations underlying \eq \eqref{e:metrst}.

Furthermore,
a running gravitational coupling constant can not be implemented in the relativistic case
as simply as in \Sec~\ref{s:denn}.
To preserve Lorentz invariance,
one has to associate this coupling with a four-dimensional scalar;
which automatically upgrades general relativity to a scalar-tensor gravity
with a non-minimally coupled scalar field.
This theory cannot be written by hand, but~it must be derived 
in a way which is similar to that used in \Sec~\ref{s:cfs}.

Comparing results of Sections~\ref{s:denn} and \ref{s:denr},
one can conclude that the definition of effective matter density is somewhat ambiguous:
in particular, it drastically depends on symmetry assumptions.
Therefore, further experimental studies should help to empirically
establish which symmetry is more appropriate to use when dealing with 
``dark''~phenomena.

\scn{Galactic Rotation Curves}{s:frc}
In this \Sec, we  demonstrate how  induced gravitational potential
can 
explain
various phenomena, which are usually  attributed
to dark matter.
Let us focus on the terms \eqref{e:grgal} and \eqref{e:grmgl}, which~were partially discussed in \Sec~\ref{s:pmgal}. 
Because they become significant at a galactic scale and above (i.e.,~a kiloparsec to megaparsec scale),
it is natural to conform them to astronomical observations; such as those of rotation curves in~galaxies.

%
In a spherically symmetric case,
velocity curves of stars orbiting 
with non-relativistic velocities
on a plane in a central gravitational potential $\grpott (\rad)$ 
can be estimated using a simple formula
$ 
v^2 
= \ror a_c = 
\ror \,
\grpott' (\ror)
$, 
where $v$ is the orbital velocity,
$a_c$ is the centripetal acceleration,
and $\ror$ is the orbit's radius.
The~cylindrically symmetric case can be considered by analogy,
by assuming various 
disk models~\cite{too63,cas83}.

Considering the terms \eqref{e:grgal} and \eqref{e:grmgl} in conjunction with the Newtonian term \eqref{e:grnew},
we thus obtain
\ba
v (\ror) 
&=&  
\left\{
\ror \Der{}{\ror} 
\left[
\grnew (\ror) + \grgal (\ror) + \grmgal (\ror)
\right]
\right\}^{1/2}
\nn\\&=&
\sqrt{
v_\text{N}^2 
+ v_\text{gal}^2 
+
\grmgal (\ror)
}
, \lb{e:ov}
\ea
where
\ba
v_\text{N}^2 
&=&
\frac{G M}{\ror}
=
\frac{a_1 q}{m \lnc} \frac{1}{\ror}
=
- \grnew (\ror)
,\\
v_\text{gal}^2 
&=&
\pwr c_b^2 =  \frac{\pwr b_0}{m} 
= \text{const}
, \lb{e:ovgal}
\ea
while the contribution from the term \eqref{e:grcos} is disregarded for now,
due the assumed smallness of the ``local'' cosmological constant
$1/\rcos^2$; 
and
the contribution from the term \eqref{e:grsmi} is disregarded due the assumed smallness of 
the corresponding characteristic length, according to  discussion in \Sec~\ref{s:psmi}.

In the case of a galaxy, 
the contribution from the Newtonian term $\grnew$ rapidly decreases as $\ror$ grows.
Correspondingly, the~main contribution would then come 
from
the second term in a row, $\grgal$, and~then from
the third term, $\grmgal$.
From \eq \eqref{e:ovgal} one can see that the contribution from $\grgal$ is constant,
which explains the average flatness of galactic rotation~curves.

Notice that the value of  velocity $v_\text{gal}$ 
depends on one of the wavefunction parameters $\pwr$
and one of quantum temperature parameters $b_0$.
Both are not \textit{a priori} fixed parameters of the model,
cf. \mbox{\eqs \eqref{e:bqtem} and \eqref{e:ava}}, 
but vary depending on the
environment and conditions:
background superfluid gets affected by the gravitational potential it induces,
because this potential acts upon the surrounding conventional matter, thus creating density inhomogeneity.
Therefore, $\pwr$ and $b_0$ should generally be different for each galaxy;
and hence should $v_\text{gal}$ be.

Similar to the case of $v_\text{gal}$,
the
parameter $a_1$ hence a value $\grmgal (\ror)$ at a fixed $\ror$
will also be dependent on the gravitating object they refer to.
This  potential 
should usually be negligible on the inner scale length of a galaxy,
but
as $R$ grows towards the extragalactic length scale, $\ror 
\gtrsim 
10 \, \text{kpc}$,
rotation curves should start to deviate from flat:
\be\lb{e:grclin}
v (\ror)
\approx 
v_\text{gal} 
\sqrt{
1+
\frac{\ror}{\lpwr}
},
\ee
which can be used for estimating the combination 
of superfluid vacuum parameters
$\pwr \lnc/a_1 $ empirically.
In cases where
the contribution from other terms of the induced potential cannot be neglected,
{Equation}
~\eqref{e:ov} must be generalized to include 
those~too.


Possible galactic-scale regions, where this non-flat asymptotics  
should become visible,
depending on a value $\lpwr$,
are the outer regions of large spiral galaxies, such as M31 or M33~\cite{rvt80,cs00,ccf09,clw10,kcc17},
where $\grmgal (r)$ can  not only overtake $\grnew (r)$ but also become
comparable with $\grgal (r)$.

\scn{Accelerating Expansion of the Universe}{s:uexp}
This phenomenon is usually explained by introducing exotic forms of relativistic matter,
such~as dark energy; usually modeled by various long-range scalar fields, which are 
assumed not to affect the numerous
particle physics experiments on the Earth.
The superfluid vacuum approach offers a simple framework,
which can explain the Universe's expansion as an observer-dependent effect,
without~involving any matter other than the background superfluid~itself.

\sscn{Conformally flat spacetime and dilaton field}{s:cfs}
Following work~\cite{z11appb},
let us consider the most simple possible special case: 
laminar flow of a logarithmic background condensate in a state
$| \wfvo \rangle$,
described by \eq \eqref{e:oF} at $b = \text{const}$,
with a constant velocity $\ve u^{(0)}$,
if viewed as,
from the F-observer's perspective, an~embedding into underlying Euclidean space.
On the other hand, what does a R-observer see?

Due to a well-known separability property
of the logarithmic Schr\"odinger equation~\mbox{\cite{bb76,af18,bcs19,lzh19,z18zna}},
the phase of
its simplest ground-state solutions is
a linear function of a
radius-vector:
\be\lb{e-phaselin}
i \ln{
\left(
\frac{\wfvo (\rve, t)}{|\wfvo (\rve, t)|}
\right)} 
\propto 
\ve u^{(0)} \cdot \rve + f (t)
,
\ee
where $\ve u^{(0)}$ 
is a constant 3-vector, and~$f(t)$ is an arbitrary function of time. 
In this case, the~fluid-Schr\"odinger analogy 
confirms
that the background condensate does flow with a constant 
velocity 
$
\ve u 
= - i \eta \vena \ln{
\left(
\wfvo 
/ |\wfvo| 
\right)
} 
\propto
\ve u^{(0)} 
$. 
Recalling \eq \eqref{e:metr1}, it means that 
the background
geometry induced by such solutions is conformally flat:
\be\lb{e-metlog}
d s^{2}_{(0)} 
\propto
\frac{1}{\dnc}
|\wfvo (\rve, t)|^2
\left[
- 
c_s^2
d t^2
+
\left(d \rve  - \ve u^{(0)} d t\right)^2
\right]
,
\ee
where
$c_s$ is given by \eq \eqref{e:cappappln},
in a leading-order approximation
with respect
to the Planck~constant.

Spacetime of a type \eqref{e-metlog} lies within a large class 
of manifolds with the
vanishing Weyl tensor -- a type $\mathbf{O}$ in the
Petrov classification. 
This is the class that all FLRW spacetimes belong to, including those which describe the Universe
with accelerating expansion -- simply written in conformally-flat coordinates
instead of comoving ones. 

Using definition \eqref{e:setdef},
we obtain
an  induced stress-energy tensor
for our system:
\ba\lb{e-setelog}
\kappa 
\widetilde{T}_{\mu\nu} 
&=&
\tilde D
\Bigl[
\nabla_\mu \nabla_\nu \dil
-
\nabla_\mu \dil \nabla_\nu \dil
\nn\\&& 
-
g_{\mu\nu}
\left(
\nabla_\lambda \nabla^\lambda \dil
+
\tfrac{1}{2}
(\tilde D-1) 
\nabla_\lambda \dil \nabla^\lambda \dil
\right)
\Bigr]\!,~~~
\ea
where $\tilde D = D -2 = 2$,
$\nabla$ is a covariant derivative
with respect to metric $g$,
and by $\dil$ 
we denote the induced scalar field:
\be\lb{e-phipsi}
\dil =
 \ln{
\left( 
|\wfvo (\rve, t)|^2/\dnc
\right)
} 
,
\ee
up to an additive~constant.

This stress-energy tensor strongly resembles 
the one occurring in
the theory of gravity with a scalar field. 
One can verify that
it
can  be  indeed derived
from the following scalar-tensor
gravity action~functional
\be\lb{e:actv}
\tilde{\cal S} 
[g,\, \not\!{\dil}]
\propto
\int d^D x
\sqrt{- g}
\,
\text{e}^{\tilde D \dil}
\left[
R + \tilde D (\tilde D + 1) 
(\nabla \dil)^2
\right]
,
\ee
where the notation ``$\not\!{\dil} $'' reminds us
that
the field $\dil$ is fixed by the solution 
of the original quantum wave equation, cf. \eq \eqref{e-phipsi},
while the variation of action must be taken with respect to the metric only.
In other words, 
both metric $g$ and dilaton $\dil$ are 
induced
by the superfluid vacuum being in a state described by $\wfvo (\rve, t)$.

Thus, we have found yet another example of the differences between the F-observer's
and R-observer's pictures of reality. 
While the former sees 
a background quantum fluid flowing with a constant velocity
in three-dimensional Euclidean space,
the latter observes itself as being inside four-dimensional spacetime governed 
by a Lorentz-covariant scalar-tensor~gravity.

An action functional \eqref{e:actv}
therefore explains why covariant models involving
scalars
provide a robust description of the large-scale
evolution of the Universe, 
agreeing with current observational data;
yet no quanta of relativistic dilaton have thus far been~detected.

This correspondence
also reveals the
limitations of the relativistic description
itself:
if the superfluid vacuum goes into a different quantum state,
then
one gets a different expression for the induced metric,
scalar, stress-energy tensor and covariant
action.
In fact, for~more complicated superfluid flows, even the condition \eqref{e-phaselin},
leading to a conformal flatness, 
can become relaxed to an asymptotic one.
Therefore, depending on the physical configuration (determined by external potential,
if any,
and boundary conditions), nonlinear coupling behaviour
and the quantum state the vacuum is in,
small fluctuations and probe particles
would obey different covariant actions.
Consequently,
a~R-observer would have to tweak its field-theoretical models by hand;
with
the unified picture being observable only
at the level
of a F-observer.

\sscn{Cosmological constant and local expansion mechanism}{s:cco}
In \Sec~\ref{s:cfs} we considered an example of the global superfluid flow
which would be seen by a relativistic observer 
as the accelerating expansion of the observable Universe.
What about locally induced gravity \eqref{e:grev}, can it cause similar effects?

Let us  consider once again the induced potentials of
Sections~\ref{s:pcos} and \ref{s:pmgal}.
The de Sitter term from \Sec~\ref{s:pcos} predominates if we consider
the physical vacuum alone, without~any generated matter therein.
This is perhaps only valid for the early Universe,
such as the one which existed during the
inflationary epoch.
In the current epoch, our spacetime can only be de Sitter asymptotically or
even
approximately;
therefore the potential $\grcos$ must be negligible, unless~one considers  
very large length scales.
For example, if~a length scale $\rcos$,
which generally depends on a massive object defining the frame of reference
of the function \eqref{e:avex},
is comparable to a size of the observable Universe 
($\sim$ 10 Gpc),
then $\rcos$ can be related to the cosmological constant $\Lambda$
as
\be\lb{e:rhlam}
\rcos^
\text{(cos)} 
= \sqrt{3/\Lambda}, \ \
\Lambda \sim 
10^{-56} \ \text{cm}^{-2}
,
\ee
therefore,
the term \eqref{e:grcos} becomes substantial only at a gigaparsec scale.
This relation yields an empirical constraint for a
combination of characteristic parameters
of superfluid vacuum, average quantum temperature and $\wfv$: 
\be\lb{e:ablam}
\frac{
a_2^
\text{(cos)} 
b_0^
\text{(cos)}
}{m  \lnc^2}
\approx
\frac{1}{6} \Lambda c^2
\sim
10^{-36} \ \text{s}^{-2}
,
\ee
see also 
the discussion around
\eq \eqref{e:bqtem}.
All this essentially means that 
$\Lambda$
is not a fundamental constant of Nature,
but a combination of various parameters of superfluid vacuum, including quantum temperature
and Gaussian width of the condensate function $\wfv$.
Therefore, its smallness, sometimes referred as ``vacuum catastrophe'' \cite{car01},
can easily be  explained by the fact that: 
either average quantum temperature across 
the Universe, or~wavefunction's width, or~both,
are sufficiently small compared to $m c^2$ and $\lnc$, respectively;
thus resulting in the overall smallness of the ratio on the left-hand side of \eq \eqref{e:ablam}.

The line element, which results from taking $\grcos$,
$\grmgal$, $\grgal$, $\grpott_0$ and $\grnew$ together,
can be written in static coordinates as
\bw
\ba
d s^2_{(\text{cos+N})}
\!&\approx&\!
- c_{(0)}^2 
\left[
1 
+ \delta_0 
- \frac{r_H}{r}
+ \beta_\pwr^2 
\ln{\!\left(
\frac{\rad}{\lnc}
\right)} 
+ \frac{r}{R_\ssmgl} - \frac{r^2}{\rcos^2}
\right] d t^2
\nn\\&&
+ \frac{d r^2}{
1 
+
\delta_0 
- r_H/r
+ 
\beta_\pwr^2 \ln{\left(\rad/\lnc\right)}+ r/R_\ssmgl - r^2 / \rcos^2
} + 
r^2 d \mesph
,
\lb{e:medsa2}
\ea
\ew
where $\delta_0 = 2 \grpott_0/c_{(0)}^2$.
In this metric,
the   
Schwarzschild term
ensures that spacetime singularity at $r = 0$ is ``dressed'' by the black hole horizon,
while at large $r$ we still have spacetime of a
FLRW type.
While from~the viewpoint of a F-observer, no spacetime singularity would
pose a problem, because~quantum wavefunction remains regular and normalizable
at any non-negative value of $\rad$.
This reaffirms our earlier statement that cosmological singularity is an artifact
of the low-momentum approximation of superfluid vacuum~\cite{z11appb,z20rss}.

According to \eq \eqref{e:memgl2},
the linear potential term \eqref{e:grmgl}, when taken alone,
induces 
the universal acceleration 
$c_{(0)}^2/2 R_\ssmgl$,
occurring due to the spatial curvature,
when seen by a R-observer in its own local static coordinate system.  
This contributes
to the Hubble expansion induced by the quadratic term \eqref{e:grcos}.
Thus,
in the relativistic picture,
the non-small fluctuation of superfluid vacuum produces an effect at the center of a gravitating
configuration;
and therefore contributes to the explicit
rotational motions of the stars inside this configuration, 
which can be seen as a consequence of curved~spacetime.

\sscn{Expansion mechanisms and cosmological coincidences}{s:coi}
Comparing \Secs \ref{s:cfs} and
\ref{s:cco},
one can see that they describe different expansion mechanisms.
The mechanism of \Sec~\ref{s:cfs} occurs due to the global flow of background superfluid, 
assumed to be laminar, which is ``seen'' by a R-observer as a FLRW-type spacetime;
the resolution of various cosmological problems 
related to this mechanism
was discussed in \Sec~5 of Ref.~\cite{z11appb}.
On the other hand,
in \Sec~\ref{s:cco}, expansion is explained as a cumulative effect from terms
in metric \eqref{e:medsa2},
which~do not vanish at spatial infinity,
induced by a ``local'' wavefunction associated with a gravitating configuration or body.
This wavefunction can be regarded as a fluctuation (not necessarily small) of the
global wavefunction from \Sec~\ref{s:cfs}.

The interplay between these mechanisms depends on the length scales of the quadratic and linear terms,
$\rcos$ and $R_\ssmgl$.
Unless 
a cumulative expansion effect from 
asymptotically non-vanishing terms taken together
is,
by some extraordinary coincidence, exactly equal
to the expansion due to the global flow mechanism, 
its rate must be different 
from that of the global flow-induced~expansion. 

The occurrence of an additional expansion mechanism,
at the scale of a supercluster, such as our Virgo or Laniakea, 
could explain the remarkable discrepancy between measurements of the Hubble constant using different methods from
those
based on the whole Universe expansion, such as cosmic microwave background radiation (CMB).
Among non-CMB methods one could mention 
Cepheid calibration, 
time-delay cosmography,
and
geometric distance measurements to megamaser-hosting galaxies~\cite{fmg01,cfs19,pbr20}. 
From a theoretical point of view,
a scenario with different expansion rates seems slightly more plausible, because~it does not require 
an explanation why accelerations from different mechanisms should be exactly equal to each other
(this coincidence should not be confused with the conventional cosmological coincidence 
which we will discuss next).

Furthermore,
our approach offers a simple explanation of the cosmological coincidence problem itself~\cite{car01}, 
in both the simplified and quantitative versions~thereof.

The simplified formulation of the cosmological coincidence states that if dark matter and dark energy were different kinds of matter, then during the Universe's evolution they 
should have evolved independently of each other;
therefore their distributions would be uncorrelated by now  - which does not seem to be the case.
Superfluid vacuum theory trivially resolves this paradox: because ``dark matter'' and ``dark energy'' are 
actually induced phenomena
and manifestations of the same object, superfluid vacuum,  
they cannot be independent from each~other.

The quantitative formulation of the coincidence problem is
an explanation requirement for why the ratio of DM- and DE-associated densities is of order one,
despite the reasons given in the simplified formulation.
In our approach, we can regard all terms of the induced
potential, which do not vanish at spatial infinity,
as being associated with ``dark'' effects;
but inside this group we cannot unambiguously separate DM-attributed effects from DE ones.
For example,
potential \eqref{e:grmgl} is intermediate
between de Sitter and logarithmic, thus it affects both galactic rotation curves and Hubble expansion,
cf. \mbox{\Secs \ref{s:frc} and \ref{s:cco}}.

If, for~simplicity,
we consider only the local expansion mechanism of \Sec~\ref{s:cco}
and omit the contribution from  $\grpott_0$,
then the cosmological coincidence can be reformulated
as the following~condition:
\be\lb{e:coi}
\frac{
\adn{gal} 
+
\adn{mgl} 
}{
\adn{mgl} 
+
\adn{dS} 
}
=
{\cal O} ( 1 )
,
\ee
where 
$\adn{} 
$'s
are average values corresponding to densities from \Sec~\ref{s:denr}.
The numerator of this ratio represents average density
of effective ``dark matter'',
while the denominator represents effective ``dark energy'' density; or at least 
the predominating  proportions thereof.

In general, this condition 
simply imposes yet another constraint for the parameters
of the theory.
It is trivially satisfied if the
involved parts of the gravitational potential, hence the associated 
densities,
are of the same order of magnitude,
if averaged on a large scale.
Moreover, even if 
$\adn{gal}$
or
$\adn{dS}$ are much smaller
than the remaining involved densities,
but the value 
$\adn{mgl}$
is substantial,
then
the
relation~\eqref{e:coi} still holds,
due to the presence of 
$\adn{mgl}$
in both parts of the ratio.

\scn{Conclusions}{s:con}
Working within the framework of the post-relativistic theory of physical vacuum,
based on the logarithmic superfluid model,
we derived induced gravitational potential, corresponding to 
a generic quantum wavefunction of the vacuum.
This mechanism is radically different from the one used in models of
relativistic classical fluids and fields, 
which are based on modifying the stress-energy tensor in Einstein field~equations.

The form of such a wavefunction is motivated by ground-state
solutions of quantum wave equations of a logarithmic type.
Such equations find
fruitful applications in the theory of strongly-interacting quantum fluids,
and have been successfully applied to laboratory  
superfluids~\mbox{\cite{z12eb,z19ijmpb,sz19}}.
We note that,
in principle, one is not precluded from adding other types of nonlinearity,
such as polynomial ones, 
into the condensate wave equations, 
but the role of  logarithmic nonlinearity is
crucial
 
Thus, we used a logarithmic superfluid model with variable nonlinear coupling,
because it accounts for an effect of the environment in a more realistic way than
the logarithmic model with a constant coupling. 
As a result,
for the trapless version of our model,
we have four parameters, but~only two of them 
are  \textit{a priori} fixed, whereas the other two 
can vary, depending on the 
quantum thermodynamic properties of 
the environment under consideration. 
Additionally, a~number of parameters come from the wavefunction solution itself.
Those are not independent parameters of the theory, but~functions thereof.
Because we do not yet know the exact form of the superfluid wavefunction,
see remarks at the end of \Sec~\ref{s:gr},
we leave those parameters to be empirically estimated, or~bound,  at~the stage
of current~knowledge. 

It turns out that  gravitational interaction has a multiple-scale structure
in our theory: induced potential is dominated by different terms at each length scale;
such that
one can distinguish  
sub-Newtonian, Newtonian (inverse-law), galactic (logarithmic-law), metagalactic (linear-law), and~cosmological (square-law) parts.
A relativistic observer,
who operates with low-momentum small-amplitude fluctuations of superfluid
vacuum, observes this induced potential by
measuring the
trajectories of probe particles
moving along
geodesics in induced four-dimensional pseudo-Riemannian spacetime.
The metric of the latter is determined by virtue of the BEC-spacetime correspondence 
and fluid-Schr\"odinger analogy, 
applied~jointly.

The sub-Newtonian part of the induced gravitational potential is
defined as  one which grows faster than the inverse law, as~distance tends towards zero.
It
can be naturally divided into the following two parts.
One part has an inverse square law behaviour,
and thus can be associated with the gravitational field caused by a $U(1)$ gauge charge,
such as an electric charge.
On a relativistic level, it is described by  Reissner-Nordstr\"om spacetime.
The other part has `inverse square times logarithm' law behaviour,
which might become substantial at both ultra-short and macroscopic distances,
depending on the values of the corresponding parameters.
If it ``survives'' at macroscopic distances, then it upgrades Newton's
gravitational constant to a function of length, such that gravity has both strong and weak~regimes.

With the potential or spacetime metric in hand,
one can, in~principle, assign effective fictitious matter density to our potential,
which corresponds to  ``dark matter'' and ``dark energy''.
This can be done in two ways:
either by Einstein field equations in a relativistic case, or~the Poisson equation in a non-relativistic one.
It should be noted that the resulting density in each of the cases can be modified, depending on whether the gravitational constant is considered to be running or not. 
This will require more verification from future experimental and theoretical studies.

Furthermore,
on a galactic scale and above, the~potential 
is dominated by non-Newtonian terms,
which do not vanish at spatial infinity. 
This explains the non-Keplerian behaviour of rotation curves in galaxies,
which is often attributed to  dark matter.
Our model, not only explains the average flatness of
galactic rotation curves,
but also makes a number of new predictions.
One of them is
the approximately linear law behavior of gravitational
potential on a metagalactic scale, which is an
intermediate scale between galactic distances and the size of the observable universe.
This should partially affect galactic rotation curves too:
as the distance from the gravitating center grows further towards the metagalactic length scale, a~squared velocity's  profile 
asymptotically changes from being
flat towards linear, cf. \eq \eqref{e:grclin}.

On the other hand,
at the largest length scale, the~induced potential displays square law behaviour.
If the quadratic term is negative-definite, then
the corresponding metric describes (asymptotically) de Sitter space,
merely written in static coordinates.
Taken together with the contribution from the linear potential term,
this explains the accelerating expansion of the corresponding spacetime region,
which~is usually associated with dark~energy.

Such expansion could supplement the ``global'' one,
caused by
laminar flow of background logarithmic superfluid absent any other matter,
which 
induces 
a FLRW-type 
spacetime.
The occurrence of more than one type of expansion mechanism,
could be responsible for the discrepancy between measurements of the Hubble constant using different~methods.

The relevant problems, such as smallness of cosmological constants and cosmological coincidence,
were also~discussed.

To conclude,
we used the BEC-spacetime correspondence and fluid-Schr\"odinger analogy
to argue that the description of reality and fundamental symmetry crucially depend on the choice of an observer. 
We demonstrated that 
both dark matter and dark energy are related phenomena, and~different manifestations of the same object, 
superfluid vacuum, which acts by 
inducing both gravitational potential and~spacetime.\\

\noindent
\textbf{Abbreviations}\\ 
The following abbreviations are used in this manuscript:\\

\noindent 
\begin{tabular}{@{}ll}
BEC & Bose-Einstein condensate\\
CDM & Cold dark matter\\
CMB & Cosmic microwave background\\
DE & Dark energy\\
DM & Dark matter\\
dS & de Sitter\\
FLRW & Friedmann-Lema\^itre-Robertson-Walker\\
F-observer & Full observer\\
RN & Reissner-Nordstr\"om\\
R-observer & Relativistic observer\\
SVT & Superfluid vacuum theory
\end{tabular}

\begin{acknowledgments}This research is supported by Department of Higher Education and Training of South Africa and in part by National Research Foundation of South Africa.
Proofreading of the manuscript by P. Stannard is greatly appreciated.
\end{acknowledgments}
\appendix
\scn{Superfluid or not}{s:sf}
When it comes to astrophysics and cosmology,
including dark matter and dark energy applications,
the notion of
superfluid is probably the most misunderstood and misused term of all.
In addition to having a catchy, but~not exactly informative, prefix,
the first thing which 
causes misunderstanding of the term is inviscid flow. 
This,
the
most striking
feature of superfluids, is often regarded as the definition thereof. 
As a result, superfluids are often confused with
perfect fluids, which
have no viscosity by definition,
and
can be easily implemented into Einstein field equations 
by virtue of a stress-energy tensor borrowed from classical fluid~mechanics.

However, perfect fluids, being classical hydrodynamical objects by construction, 
can not properly reflect the essentially quantum nature of the superfluidic matter
that
we know of.
Therefore, perfect or non-quantum inviscid fluids can be used as a crude approximation, at~best,
of superfluids.

Additionally, superfluids are often also confused 
with the concomitant phenomenon of Bose-Einstein condensation (BEC),
which is
another kind of quantum matter occurring in low-temperature condensed matter, 
such as cold cesium atoms in a trap. 
However, in~the superfluid helium phase, for~example, the~BEC's content comprises only about ten per cent.
Therefore, 
this~condensate alone cannot 
fully account for dissipation-free flow and the other distinctive features of superfluids.
Even though Bose-Einstein condensates do share certain features with superfluids, 
this~does not imply that they are 
superfluidic in~general.

In particular, quantum excitations in laboratory superfluids are
known to have dispersion relations of a distinctive shape
called the Landau ``roton'' spectrum.
This shape of the spectral curve is crucial, as~it ensures 
the suppression of dissipative fluctuations at a quantum level~\cite{z12eb}.
If plotted as an excitation energy versus momentum,
the curve starts from the origin, climbs up to a local maximum (called the \textit{maxon} peak), 
then slightly descends to a local nontrivial minimum (called the ``roton'' energy gap);  and then grows  again, this time all the way up, 
to the boundary of the theory's applicability range, cf. a solid curve in the Figure~1a from \Ref~\cite{z20ijmpa}.
In fact it is not the roton energy gap alone, but~the energy barrier formed by the maxon peak and roton minimum in momentum space,
which ensures the above-mentioned suppression of quantum fluctuations in quantum liquid
and, ultimately, causes its flow to become inviscid.
In other words, it is the global characteristics of the curve, not just the existence of a nontrivial
local minimum and related energy gap,
which are required for superfluidity to occur. 
Obviously, Landau's shape is a non-trivial property
which cannot possibly occur in all quantum~liquids.

The final reason for the misuse of the term `superfluid'
is the extensive, but~not always careful, utilization of relativistic scalar field models 
in astrophysics and cosmology.
Historically, the~four-dimensional scalar field came about
as a bold
extrapolation of a non-relativistic wavefunction 
into the realm of Lorentz-symmetric theories.
However, when it comes to quantum liquids and condensates,
there is a fundamental difference between relativistic scalar field and condensate wavefunction,
which makes a correspondence between them far from~isomorphic.

The
fluid
condensate wavefunction 
obeys both a normalization condition \eqref{e:norm} and a wave Equation~\eqref{e:becgeneq},
therefore, it
is a three-dimensional Euclidean scalar related to a ray in the associated
Hilbert space.
One can see that \eqs \eqref{e:norm}--\eqref{e:ftlan}
are essentially non-relativistic and three-dimensional.
While one can still make the last
two 
relativistic, by~replacing 
derivative parts with the Lorentz-covariant 
analogues thereof,
\eq \eqref{e:norm} strongly violates the Lorentz invariance.
This~condition requires the foliation of a spacetime manifold into $(3+1)$-dimensional spacelike hypersurfaces,
and
makes mass-energy a three-dimensional scalar, not a time component 
of a four-dimensional vector.

As a result, relativistic scalar fields models offer a useful, but~approximate description 
of superfluidic phenomena (which is valid for small wavefunction amplitudes and low momenta of
excitations, running ahead),
while the rigorous extrapolation of BEC and superfluid notions into
the relativistic domain
requires special treatment.
Therefore, it is not surprising that some ``superfluid'' models of 
dark matter; which are 
based on classical perfect fluid models, scalar field theories
or scalar-tensor theories of gravity,
turn out to be vulnerable to experimental verification~\cite{lmo19}.

\scn{Assumptions and approximations}{s:aa}
Let us summarize and enumerate the main 
simplifying assumptions and approximations underlying
the derivation of our gravitational potential.
We used them to keep our calculations as analytical and non-perturbative
as possible, which is crucial for an essentially nonlinear theory, such as~ours.

First, 
it should be emphasized that superfluid vacuum theory, even when  narrowed down to its logarithmic version, is a framework which potentially contains a set of models.
Our chosen underlying quantum superfluid model, defined by \eq \eqref{e:vcm},
is a minimal one.
It can easily be  expanded by adding polynomial terms $|\Psi|^k$
to the logarithmic nonlinearity in wave equations,
to make it describe the phenomena in a more precise way.
The reason for this is that the logarithmic nonlinearity is a leading-order approximation for
the condensate-like matter, as~discussed in \Ref~\cite{z18zna};
but non-logarithmic terms can also come into play: one example is to be found in \Ref~\cite{sz19}.
The nonlinear coupling function $b = b (\rve, t)$ can also be made more detailed,
to account for the thermodynamic environment in a more realistic~way.

Second, for~a chosen model, the~signs of coupling parameters can be changed, which is often equivalent 
to changing the topological sector a model belongs to.
Because we do not precisely know the topological structure of the physical vacuum we live in, it
essentially doubles the number of candidate models to be tested empirically.
For example, changing the sign of the field-theoretical potential;
or, alternatively, the~overall sign of the nonlinear coupling $b$;
switches between the topologically trivial, which is considered in this paper, and~
the topologically non-trivial sectors.
In a logarithmic liquid model with constant nonlinear coupling, switching to the topologically non-trivial sector 
of the theory means
that wavefunction changes from a droplet-like non-topological soliton type to a bubble-like 
topological soliton type, 
as discussed in \Refs~\cite{z18epl,z19ijmpb}. 
Consequently, the~forms of a trial wavefunction and induced potential
in \Sec~\ref{s:gr} would also~change.

Third,
we have avoided the problem of ambiguity in choosing trapping potential and 
boundary conditions by making the former identically zero, and~the latter 
of a conventional quantum-mechanical type on an infinite spatial domain
of trivial topology $\mathbb{R}^3$.
These are simplifying assumptions which are yet to be proven to work,
otherwise they must be replaced with something more~sophisticated.

Fourth, even if we have chosen our model correctly, there still remains the ambiguity of
how to determine the state of the vacuum---is it in a ground state, an~excited but metastable state, a~pure or mixed state,
a superposition of states, 
or even in a quantum transition between states?
Any~new empirical information about this could drastically change the assumptions
underlying the computations in \Sec~\ref{s:gr}.

Fifth, the~technical approximation
which led us from \eqs \eqref{e:avex}--\eqref{e:ava}
might 
oversimplify the picture.
It is suitable for the purposes of this study,
but should be modified in more precise~considerations.

Finally, throughout the paper, we consider the physical vacuum alone:
assuming that its small excitations,
which would be observed by a R-observer as relativistic matter with deformed 
dispersion relations~\cite{z20ijmpa,z11pla,sch17,csa18,ol19,plj20},
do not back-react, for~example, via interaction between scalar and tensor modes.
This is obviously an over-simplified picture of reality,
which is
sufficient for our current purpose, but~unlikely to be valid in~general.

\end{document}